\documentclass[a4paper,11pt]{article}
\pdfoutput=1 

\usepackage{jinstpub} 
\usepackage{multirow}

\title{\boldmath Developments and improvements of radiopure ZnWO$_{4}$ anisotropic scintillators}

\author[a,b]{P.~Belli}
\author[a,b]{R.~Bernabei}
\author[c,d]{F.~Cappella}
\author[a,b,e]{V.~Caracciolo}
\author[a,b]{R.~Cerulli}
\author[f]{N.~Cherubini}
\author[g]{F.~A.~Danevich}
\author[c,d]{A.~Incicchitti}
\author[g]{D.~V.~Kasperovych}
\author[a,b]{V.~Merlo}
\author[f]{E.~Piccinelli}
\author[g]{O.~G.~Polischuk}
\author[g]{V.~I.~Tretyak}

\affiliation[a]{INFN sezione Roma ``Tor Vergata'', I-00133 Rome, Italy}
\affiliation[b]{Dipartimento di Fisica, Universit\`a di Roma ``Tor Vergata'', 
	I-00133, Rome, Italy}
\affiliation[c]{INFN sezione Roma, I-00185 Rome, Italy}
\affiliation[d]{Dipartimento di Fisica, Universit\`a di Roma ``La Sapienza'', I-00185 
	Rome, Italy}
\affiliation[e]{INFN, Laboratori Nazionali del Gran Sasso, I-67100 Assergi (AQ), 
	Italy}
\affiliation[f]{Enea, Italian National Agency for New Technologies, Energy and 
Sustainable Economic Development, C.R: Casaccia, Roma 00123, Italy}
\affiliation[g]{Institute for Nuclear Research, 03028 Kyiv, Ukraine}

\emailAdd{riccardo.cerulli@roma2.infn.it}

\abstract{
The ZnWO$_4$ is an anisotropic crystal scintillator; for its peculiar characteristics,
it is a very promising detector to exploit the so-called directionality approach in the investigation 
of those Dark Matter (DM) candidates inducing nuclear recoils. 
Recently, in the framework of the ADAMO project, an R\&D to develop high quality and ultra-radiopure 
ZnWO$_4$ crystal scintillators has been carried out. In the present paper the
measurements to study the anisotropic response of a ZnWO$_4$ to $\alpha$ particles and to 
nuclear recoils induced by neutron scattering are reported.
Monochromatic neutrons have been produced by a neutron generator at ENEA-CASACCIA. 
The quenching factor values for nuclear recoils along different crystallographic 
axes have been determined for three different nuclear recoils energies. 
These results open the possibility to realize a pioneer 
experiment to investigate the above mentioned DM candidates by means of the directionality.
}


\proceeding{N$^{\text{th}}$ Workshop on X\\
  when\\
  where}

\begin{document}
\maketitle
\flushbottom

\section{Introduction}

The presence of Dark Matter (DM) candidates inducing nuclear recoils can be investigated
by exploiting the so called directionality approach based on the correlation 
between the direction of nuclear recoils in the 
target detector and the impinging direction of the DM wind in the Galactic halo.
This approach can be complementary to the model independent annual modulation signature 
that has been successfully exploited by the 
DAMA/NaI, DAMA/LIBRA-phase1 and DAMA/LIBRA-phase2 experiments giving 
a positive model independent result with high confidence level \cite{RNC,modlibra,modlibra2,modlibra3,npae18,modep19}.

The idea to use anisotropic crystal scintillators to pursue the directionality approach 
was proposed in Ref. \cite{direz1, direz2}.
The light output of the anisotropic scintillators for heavy particles ({\it p}, $\alpha$, nuclear 
recoils) depends on the direction of the particles with respect to the crystal axes.
This offers a possibility to point out the presence of heavy ionizing particles with 
a preferred direction (such as recoil nuclei induced by DM candidates interaction) 
with respect to the electromagnetic background by comparing the low energy 
distributions measured using different orientations of the crystal axes 
during the day \cite{direz1,direz2,direz3}. 
The ZnWO$_4$ is a very promising crystal  for this purpose
offering suitable features \cite{Danev05,direz3,zwo1,zwo2,zwo3,zwo4,Cer17}. 
This motivated the proposal of ADAMO (Anisotropic detectors for 
DArk Matter Observation), a project to perform an R\&D to develop 
low background, high quality ZnWO$_4$ crystals for a pioneering experiment to investigate the directionality \cite{direz3,Cer17,zwoconf9,zwoconf8,zwoconf5}.
In this paper the recent results of the R\&Ds will be briefly presented with 
particular regard to the recent measurements performed with a neutron 
generator at ENEA-CASACCIA \cite{neut19}.

\section{ZnWO$_4$ anisotropic scintillator}

In the last years several ZnWO$_4$ detectors have been developed
in the framework of the collaboration between the DAMA
group of INFN and the INR-Kyiv group \cite{zwo1,zwo2,Belli08,Belli09}. Formerly, some crystals were produced  
by the Institute for Scintillation Materials (ISMA, Kharkiv, Ukraine) and, later on,
a collaboration
with the  Nikolaev Institute of Inorganic Chemistry (Novosibirsk, Russia) has started. In this last Collaboration 
a R\&D to produce ultra-radiopure ZnWO$_4$ by using the 
low-thermal gradient Czochralski technique in a platinum crucible 
is ongoing \cite{Gala09}. 
The produced crystals have been deployed in the
underground facility DAMA/R\&D at the Gran Sasso laboratory 
(LNGS) \cite{Belli08,Belli09,zwo1,zwo2}.
Measurements and R$\&$D studies have shown  the competitiveness of ZnWO$_4$ scintillators 
for a DM experiment based on directionality. In fact, 
the light output and the time profile of the scintillation pulse for heavy 
particles depends on the direction of such particles with respect to the crystal
axes while no difference is observed for $\gamma/\beta$ radiation \cite{Danev05}.
The shape of the scintillation pulse is also 
different for $\gamma(\beta)$ radiation and $\alpha$. 
This pulse shape discrimination capability
can potentially be  of interest not only for a DM experiment but also for double beta decay searches.
The ZnWO$_4$ offers also a high atomic weight and the
possibility to realize single crystals with masses of some kg \cite{Gala09}. 
Moreover, the presence of three target nuclei
with very different masses (Zn, W and O) makes these scintillators sensitive
to both small and large mass DM candidates, as also the NaI(Tl) is.
The recently developed  ZnWO$_4$ scintillators have very good level of radiopurity. 
The measured upper limits are: $20$ $\mu$Bq/kg for the $^{40}$K, 
 2 $\mu$Bq/kg for the  $^{226}$Ra and in the range (0.17-1.3) $\mu$Bq/kg
for the $^{228}$Th \cite{zwo1,zwo4}.

A further radio-purification of ZnWO$_{4}$ crystal scintillators is feasible. 
The R\&D is still ongoing. 

As confirmed by the measurements performed at LNGS, the crystals 
have relatively high light output at room temperature, being about 
20\% of the Na(Tl) scintillator.
It appears feasible to improve the light output of the crystal 
considering that it can increase when working at low temperature \cite{Nag09}.
To study this feature, a small cryostat is currently under test at LNGS 
to implement and optimize the cooling system; 
this system will allow to reach a stable working temperature around -50$^{\circ}$C. 

In order to employ the ZnWO$_{4}$ for DM investigation with the directionality approach,
it is crucial to measure and quantify the anisotropy of the scintillator for nuclear recoils.
For this purpose a campaign of
measurements has been recently performed by using a  7.99 g mass ZnWO$_4$ crystal scintillator 
of ($10\times10\times10.4$) mm$^3$, in the framework of the ADAMO project \cite{neut19}.
The crystal has been obtained by a second crystallization procedure using low-thermal 
gradient Czochralski technique from zinc tungstate crystals made from tungsten oxide additionally purified 
by double sublimation of tungsten chlorides \cite{Shl2017}.
The crystallographic axes were identified by the producer and experimentally verified.

Before measuring the response of the ZnWO$_4$ crystal with a monochromatic neutron 
source at ENEA-CASACCIA, the crystal has been irradiated with $\alpha$ particles at LNGS.
In the following the obtained results are briefly reported, while more details can be found in \cite{neut19}.

\section{Measurements with $\alpha$ particles}

The  ZnWO$_4$ crystal ($10\times10\times10.4$ mm$^3$) has been coupled to a Hamamatsu H11934-200 PMT 
(Ultra Bialkali photocathode with an effective area of $23 \times 23$ mm$^2$ and
quantum efficiency $\simeq 43\%$ at 400 nm and $\simeq  25\%$ at  500 nm).
The scintillation profiles coming from the detector have been recorded by 
a LeCroy WaveSurf24X-sA oscilloscope (4 chn, 2.5 GSamples/s, 200 MHz) 
in a time window of 100 $\mu$s.
The measurements have been performed by using an $^{241}$Am source and
various sets of thin mylar films as absorbers to decrease the $\alpha$ particles energy.
The beam of $\alpha$ particles has been collimated before reaching the crystal face. 
The energies of $\alpha$ beam have been measured with 
a CANBERRA Alpha Spectrometer (model 7401VR).
The energy scale of the crystal for each measurement has been calibrated by
using $^{137}$Cs and $^{22}$Na $\gamma$ sources.

\begin{figure}[!ht]
	\centering
	\vspace{-0.4cm}
	\includegraphics[width=0.55\textwidth]{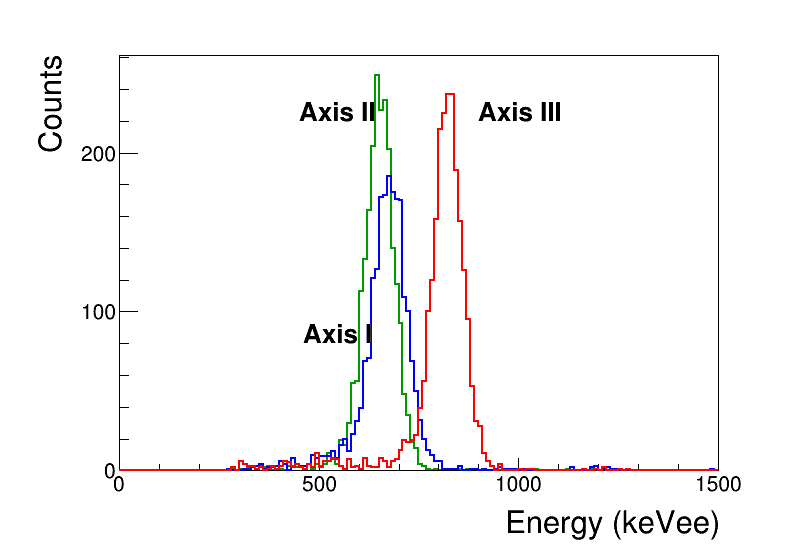}
	\vspace{-0.4cm}
	\caption{\footnotesize{Energy spectra of 4.63 MeV $\alpha$ particles impinging along the three axes of the crystal.
		}}
		\label{fg:alfa1}
	\end{figure}

\begin{figure}[!ht]
	\centering
	\vspace{-0.4cm}
	\includegraphics[width=0.65\textwidth]{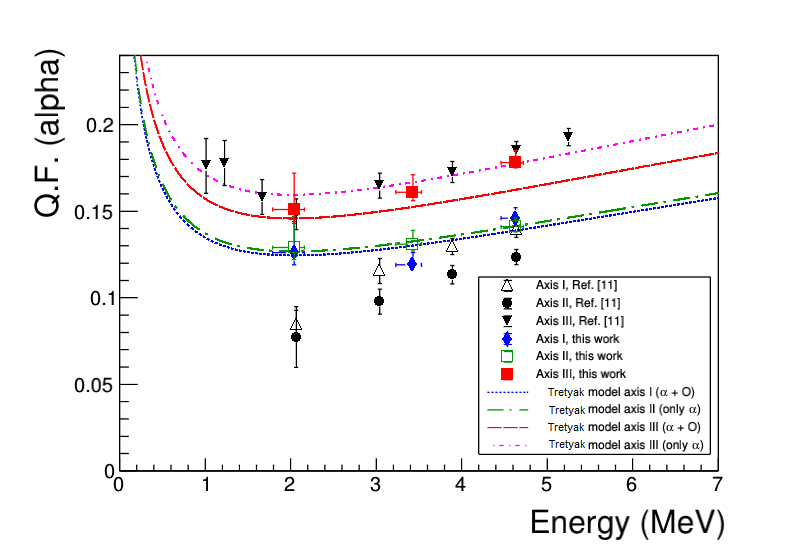}
	\vspace{-0.4cm}
	\caption{\footnotesize{Dependence of the $\alpha/\beta$ ratio on the energy of the $\alpha$ particles measured with a ZnWO$_4$ scintillator in Ref. \cite{Danev05} (black points) compared with those reported in this paper (colored points). 
	The anisotropic behavior of the crystal is evident.
	The models for each crystal axis, obtained 
	following the prescription of Ref. \cite{Tret2010} are also reported. The curves have been obtained 
	by fitting the data from $\alpha$'s and from recoils or only from $\alpha$'s.
		}}
		\label{fg:alfa2}
	\end{figure}

The typical energy distributions of the $\alpha$ particles impinging along 
the three axes of the crystal are shown in Fig. \ref{fg:alfa1}.
The ZnWO$_4$ crystal was irradiated in the directions perpendicular to 
the  (100), (001) and (010) crystal planes: hereafter crystal axes 
I (blue on-line), II (green on-line) and III (red on-line), respectively in Fig. \ref{fg:alfa1} and in Fig. \ref{fg:alfa2}.

In Figure \ref{fg:alfa2} the dependence of 
the $\alpha/\beta$ ratio\footnote{$\alpha/\beta$ ratio (for $\alpha$ particles) and 
quenching factor (for ions in general) are determined as the ratio of particle's energy measured 
in scale calibrated with $\beta$ or $\gamma$ sources to the real energy of the particle.}
as a function of energy for the three different directions of the $\alpha$ beam relatively to the crystal planes is shown.
In particular, the quenching factor for $\alpha$ particles
measured along the crystal axis III is about 1.2 times larger than
that measured along the crystal axes I and II. 
Instead, the quenching factors measured along the crystal axes I and II are quite similar. 
The error bars are 
mainly due to the uncertainty of the alpha energy which is slightly degraded in air.

The quenching factors and the anisotropic effect reported here are in reasonable agreement with those of Ref. \cite{Danev05},  as shown in Fig. \ref{fg:alfa2}; in the figure the behavior of the $\alpha/\beta$ ratio as expected for each crystal axis in the model of Ref. \cite{Tret2010} is also reported. The curves have been obtained 
by fitting the data from $\alpha$'s and from recoils or only from $\alpha$'s.

Therefore, the data confirm the anisotropic features of ZnWO$_4$ crystal scintillator in case 
of $\alpha$ particles.

\section{Measurements with neutrons}

Monochromatic neutrons have been generated by the Thermo Scientific portable generator MP 320. 
Neutrons are produced in the $d(t,\alpha)n$ reaction with energy around 14.7 MeV by
accelerating deuterons toward a tritium target in electric potential.
For the requirements of the experiment a configuration with
beam  acceleration voltage of 60 kV and a beam current of 40 $\mu$A
have been adopted in order to maximize the neutron yield 
and the stability of the beam operation; the production rate 
was around 10$^7$ n/s. In the experimental setup, neutrons leaving the target 
at 90 degrees to the forward direction have been used; at this angle, 
simple kinematics predicts a value $E_n=14.05$ MeV 
for a beam acceleration voltage of 60 kV. 

In a single elastic scattering of neutrons with target nuclei, nuclear recoils are induced, 
and the scattered neutron is detected by two neutron detectors placed 
at a given scattering angle. 
In such a configuration the energy and the direction of the recoiling nucleus is fixed
and, by measuring the energy released in the ZnWO$_4$ detector 
-- in keV electron equivalent (keVee) -- it is possible to determine the quenching factor.
By changing the crystal axes orientation, the quenching factor for the different axes 
can be measured.

\begin{figure}[!th]
	\begin{center}
		\includegraphics [width=0.5\textwidth]{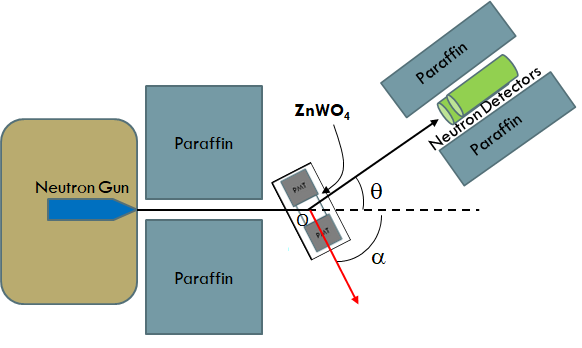}
	\end{center}
	\vspace{-0.8cm}
	\caption{\footnotesize{
			Schematic view from the top of the experimental set-up. The two neutron detectors are one above the other. 
	}}
	\label{fg:schema}   
\end{figure}

A scheme of the set-up is shown in Fig. \ref{fg:schema}.
\normalsize
The ZnWO$_4$ detector is placed in front of the neutron channel on a 
revolving platform around a vertical axis ($O$ axis),
allowing us to fix the direction of the crystal axis with respect to 
the impinging neutrons: the $\alpha$ angle as depicted in Fig. \ref{fg:schema}.
The ZnWO$_4$ crystal is centered on the $O$ axis and is optically coupled to two 
Hamamatsu H11934-200 PMTs on opposite faces.
In the set-up, the PMTs are on the faces perpendicular to the crystal axis 
III; thus, $\alpha$ identifies the angle between the
crystal axis III and the impinging neutrons direction. The crystal axis I 
is also on the horizontal plane, while 
the axis II is vertical. The ZnWO$_4$ crystal and the two PMTs are inside a 
black plastic box.
The neutrons scattered off the ZnWO$_4$ crystal are tagged by two neutron 
detectors by Scionix employing EJ-309 liquid scintillator. 
The EJ-309 scintillator has a very high 
capability to discriminate neutrons interactions 
from the gamma background by Pulse Shape Discrimination (PSD).
The neutron detectors are placed 82 cm far from $O$ axis, one above the 
other to maintain the same neutron scattering angle $\theta$ (see Fig. 
\ref{fg:schema}) and to improve the solid angle acceptance.
They are held by an arm and free to rotate around the $O$ axis.

The signals from the two PMTs coupled to the ZnWO$_4$ detector are summed
and recorded by a CAEN DT5720 transient digitizer with 250 MS/s sample rate.
The trigger is obtained by the coincidence between a signal in the 
ZnWO$_4$ detector and in the EJ-309 detectors within a time window of $\pm 500$ ns.

The energy detected by the ZnWO$_4$ detector is evaluated from the digitized pulse area.
$^{133}$Ba and $^{137}$Cs sources were used to calibrate the detector energy scale.
The typical energy resolution was $\sigma/E = 4.4\%$ at the 662 keV $\gamma$ peak.
The energy calibration has been performed before and after neutron irradiation.

\begin{figure}[!th]
	\begin{center}
		\includegraphics [width=0.49\textwidth]{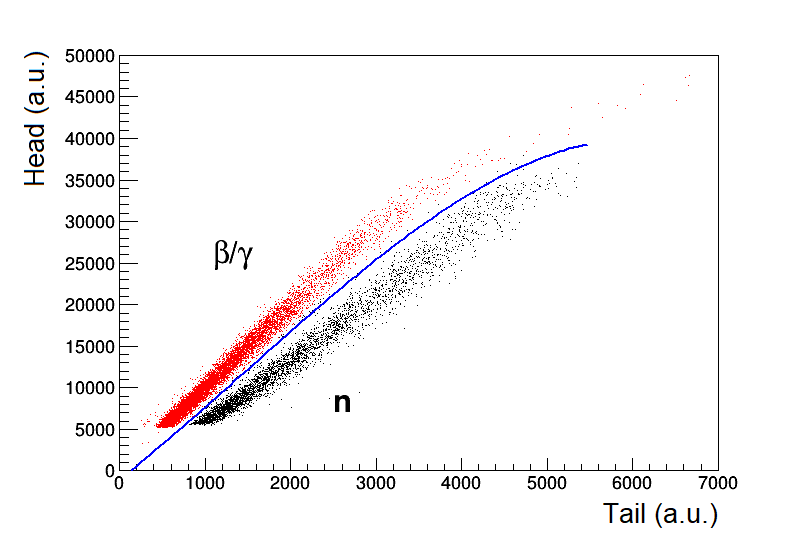}
		\includegraphics [width=0.49\textwidth]{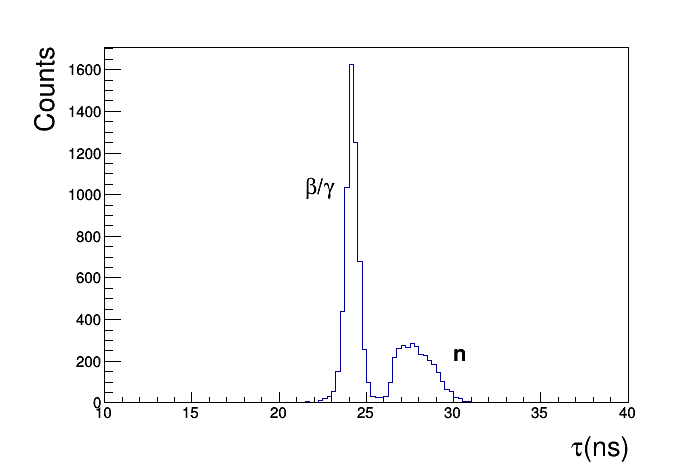}
	\end{center}
	\vspace{-0.8cm}
	\caption{\footnotesize{
			Example of gamma/neutron separation by PSD in EJ-309 liquid scintillator. Left: head/tail 
			analysis. Right: distribution of the mean time, $\tau$, variable. 
	}}
	\label{fg:psd}   
\end{figure}

Neutron events in the EJ-309 detectors have been selected by a PSD data analysis based on: 
i) head/tail analysis; ii) analysis of the mean time, $\tau$, of the time profile of 
the pulse. A typical example of the separation between gamma and neutrons exploiting
the two techniques is reported in Fig. \ref{fg:psd}-left and in Fig. \ref{fg:psd}-right, respectively.

To select the nuclear recoils induced by the elastic scattering of neutrons 
over  the background, an important quantity is the time passing between the signals in the ZnWO$_4$ detector
and the neutron detectors EJ-309 (Time of Flight, $TOF$). 
A variable $\Delta t = t_0^{EJ} - t_0^{ZWO}$, can be defined 
where $t_0^{EJ}$ ($t_0^{ZWO}$) is the starting time of the EJ-309 (ZnWO$_4$) pulse.
The transit time of the neutron detectors PMT is $\approx 40$ ns 
while it is $\approx 6$ ns for the Hamamatsu H11934-200.
The $\Delta t$ variable is shifted with respect to $TOF$ and it has been calibrated considering the coincidences between high energy events in the ZnWO$_4$ detector
and gamma events in the neutron detectors. An interval $\Delta t \approx 34$ ns has been obtained
from the data; it is in agreement with the expectation. This $TOF$ calibration has been taken into account in the analysis.

For a fixed scattering angle $\theta$ and crystal axis, the event pattern searched for
was represented by a scintillation pulse in the ZnWO$_4$ detector in coincidence with a neutron in the EJ-309 detectors. 

\begin{figure}[!t]
	\begin{center}
		\includegraphics [width=0.4\textwidth]{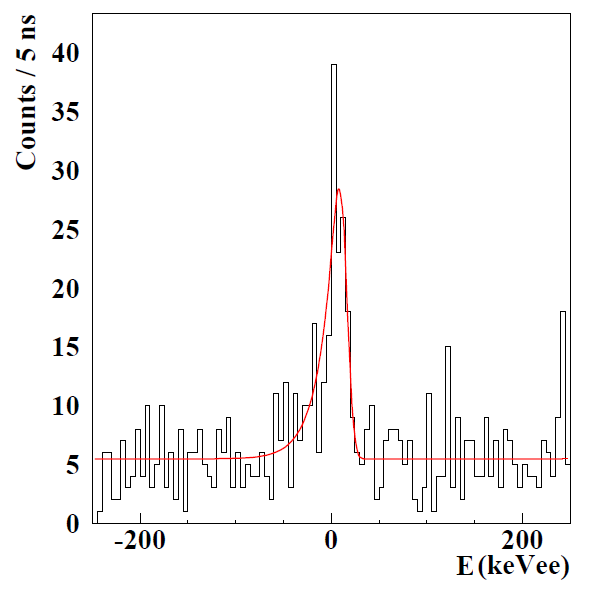}
		\includegraphics [width=0.49\textwidth]{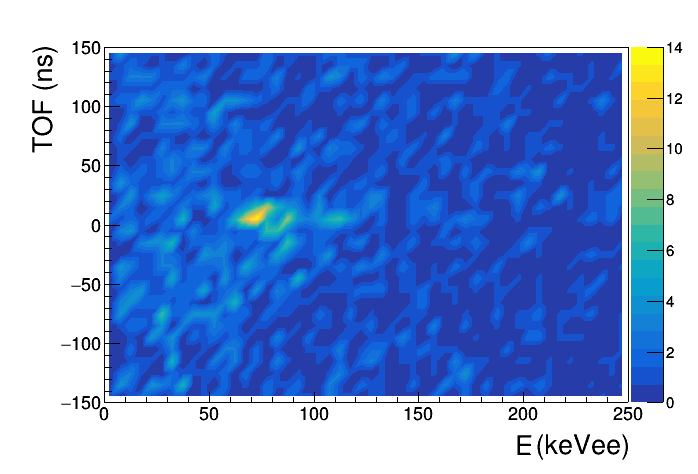}
	\end{center}
	\vspace{-0.8cm}
	\caption{\footnotesize{
			$TOF$ distribution and bi-dimensional plot $TOF$ vs ZnWO$_4$ energy (in keVee) 
			for coincidences obtained after selecting neutrons in the neutron detectors,
			for the case of scattering angle $\theta=70^{\circ}$ and axis I.
			Left: The $TOF$ distribution shows a continuum due to random coincidences and a clear peak, 
			that is in agreement with the expected $TOF$ for neutrons after elastic 
			scattering off the nuclei of ZnWO$_4$ detector. The peak shows a tail on the left part
			due to the first photoelectron delay in ZnWO$_4$. The fit with the 
			function of eq. \ref{eq:func} is superimposed (see text).
			Right: A clear peak is present at the proper value of $TOF$ and at the  given energy. 
			The peak position in the distribution measured by the ZnWO$_4$ detector provides the 
			quenching factor of the O nucleus for the scattering-angle and crystal-axis used.
	}}
	\label{fg:tof_n}   
\end{figure}

In Fig. \ref{fg:tof_n}-left the $TOF$ distribution is depicted for the case of 
$\theta=70^{\circ}$ and crystal axis I, while in Fig. \ref{fg:tof_n}-right 
is shown the bi-dimensional plot $TOF$ vs ZnWO$_4$ energy ($E$ is in keVee).
These plots show a continuum due to random coincidences, a clear excess of events
with energy around 80 keVee, and $TOF$
in agreement with the expectation for scattered neutrons on target nuclei.
The peak in the $TOF$ variable shows a tail on the left
due to the first photoelectron delay in ZnWO$_4$ (effective average scintillation decay time $\approx$ 24 $\mu$s).
The observed excess can be ascribed to  the O recoils in the ZnWO$_4$ detector 
and its position provides the quenching factor of this nucleus for the scattering-angle and crystal-axis used. The Zn and W recoils are expected to be well below the energy threshold that has been used in the present experimental conditions.

The behavior of the $TOF$ distribution can be explained by considering that events belong to the sum of two main contributions:
1) one due to the coincidences caused by the elastic scatterings of neutrons 
on the oxygen nuclei,
and 2) the random flat coincidences ($f_{rnd}$). To build a model for the $TOF$ behavior, some
assumptions can be considered: i) the signal of the neutron detectors is ``prompt" (that is the time delay between 
the neutron interaction and the starting time of the EJ-309 pulse is mostly equal to the PMT transit time); 
ii) the ZnWO$_4$ time delay between the neutron interaction and the starting time of the pulse is given 
by the PMT transit time plus the delay of the first photoelectron in the ZnWO$_4$ detector 
(effective average scintillation decay time $\approx$ 24 $\mu$s \cite{Danev05}); 
iii) the probability to have the first photoelectron in ZnWO$_4$ at time $t$ is $\frac{1}{\tau}e^{-t/\tau}$,
where $\tau$ is
a free parameter linked to the effective scintillation decay time and the number of available photoelectrons;
iv) the fluctuations of the transit times of the PMTs 
of the ZnWO$_4$ and of EJ-309 detectors, and the time resolution 
of the  digitizer (2 ns bin size for the neutron detectors and 4 ns 
for the ZnWO$_4$ detector) are taken into account by making a convolution 
of the time distribution with a Gaussian function having a characteristic spread $\sigma$.
Thus, the model function of $TOF$ can be written as:

\begin{equation}
f(TOF | f_{rnd}, A, \tau, \sigma, TOF_0) = f_{rnd} + A\int_0^\infty e^{-x/\tau} e^{-    \frac{  (x + TOF - TOF_0 )^2}{2\sigma^2}               }  dx
\end{equation}

\noindent where $TOF_0$ represents the expectation value of the neutrons time of flight
and $A$ is a normalization factor. By solving the integral one gets:

\begin{equation}
\label{eq:func}
f(TOF) = f_{rnd} + B e^{-\frac{TOF_0-TOF}{\tau}} \left[1 - erf  \left(
-\frac{TOF_0-TOF}{\sqrt{2}\sigma} + \frac{\sigma}{\sqrt{2}\tau} \right) \right]
\end{equation}

\noindent where $B$ includes only constant factors: $B = A \sqrt{\frac{\pi}{2}} \sigma e^{\frac{\sigma^2}{2\tau^2}}$.
The result of the  $TOF$ distribution fits with the function given in eq. \ref{eq:func} is plotted in Fig. \ref{fg:tof_n}-left; it gives:
$f_{rnd}  \approx 5.5 $ counts/ 5 ns, 
$B  \approx  21.6 $ counts/ 5 ns, 
$\tau  \approx 16.9$ ns, $\sigma \approx 5.8$ ns and 
$TOF_0 \approx  15.4$ ns. The model reproduces well the experimental data and is
in good agreement with the expectations. A Monte Carlo simulation gives similar results.

\begin{figure}[!t]
	\begin{center}
		\includegraphics [width=0.49\textwidth]{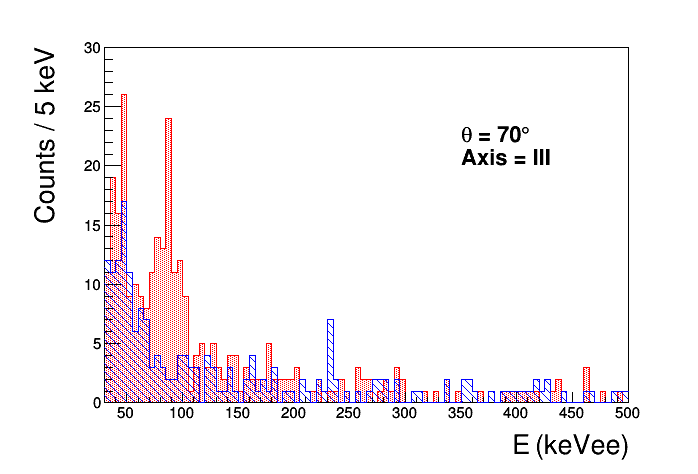}
		\includegraphics [width=0.49\textwidth]{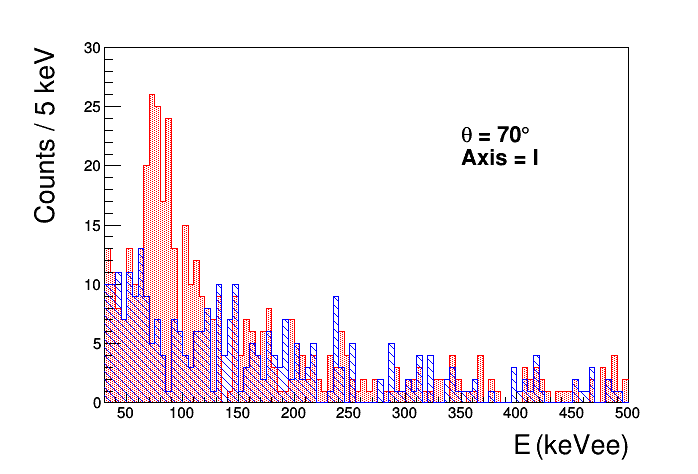}
	\end{center}
	\vspace{-0.8cm}
	\caption{\footnotesize{
			Energy spectra measured by the  ZnWO$_{4}$ detector (in $\gamma$ scale, keVee) 
			for the case of $\theta = 70^{\circ}$ and axes III and I.
			Only events identified as neutrons in the neutron detectors are selected.
			Light (red on-line) histogram: events selected in the proper window of the $TOF$ variable
			($-20$ ns $< TOF < 30$ ns).
			Dashed (blue on-line) histogram: events selected in the off-window ($60$ ns $< TOF < 110$ ns).
	}}
	\label{fg:ezwo}   
\end{figure}

For the estimation of the quenching factors three scattering angles were considered. 
Since the responses of the crystal axes I and II are rather
similar, only the case of axis I was considered.
Figure \ref{fg:ezwo} shows an example of the energy
distributions of the detected energy, $E$, in the ZnWO$_4$ detector, measured in keVee, ($\theta = 70^{\circ}$ and axes III and I).
Two distributions are reported for each plot: one is obtained by selecting events in the time window 
expected for the $TOF$ of neutrons elastically scattered off the nuclei of ZnWO$_4$ detector,
$-20$ ns $< TOF < 30$ ns (light red histogram); the other is presented by selecting  
off-window events (random coincidences), $60$ ns $< TOF < 110$ ns (blue histogram).
The off-window events are related to random coincidences and, therefore, their distribution
is the background distribution in the histogram of the in-window events.
The peaks are evident and can be ascribed to the oxygen nuclear recoils.
The positions of the peaks are obtained by fitting them with a Gaussian curve plus an exponential function,
that simulates the background of the random coincidences.
The energy resolutions, $\sigma$, of the peaks are between 8 and 12 keVee; the values 
are well in agreement with the energy dependence: $\sigma \propto \sqrt{E}$.

\begin{table}[!ht]
	\caption{\footnotesize{Summary table of the peak position due to oxygen nuclear recoils.
			For each scattering angle, $\theta$, and for the different axes of the ZnWO$_4$ crystal, the peak position, $E$, the energy resolution, $\sigma$,
			the expected recoil energies for the oxygen nucleus, $E_{R,O}$, 
			the quenching factors, $Q$, and the anisotropy, $Q_{III}/Q_{I}$ ratio, 
			are reported.}}
	\begin{center}
		\resizebox{0.7\textwidth}{!}{ 
			\begin{tabular}{c|c|c|c|r|c|c}\hline
				Scattering angle, & Crystal  & $E$       &  $\sigma$  &    \multicolumn{1}{c|}{$E_{R,O}$}       &  Quenching            &  $Q_{III}/Q_{I}$                   \\
				$\theta$               & axis       & (keVee)              &  (keVee)   &    \multicolumn{1}{c|}{(keV)}           &  factor, $Q$          &                                    \\		
				\hline\hline
				\multirow{2}{*}{80$^{\circ}$}     & III      &  $99.3\pm2.5$   & \,\;9    & \multirow{2}{*}{1402}    &  $0.0708\pm0.0018$    & \multirow{2}{*}{$1.174\pm0.051$}  \\
				& I        &  $84.5\pm2.9$   &   12    &                                      &  $0.0603\pm0.0021$    &   \\
				&    &  & &     &  &   \\
				\multirow{2}{*}{70$^{\circ}$}     & III      &  $86.5\pm2.0$   & \,\;7    & \multirow{2}{*}{1128}    &  $0.0767\pm0.0018$    & \multirow{2}{*}{$1.121\pm0.038$}  \\
				& I        &  $77.2\pm1.9$   &   10    &                                      &  $0.0684\pm0.0017$    &   \\
				&    &  & &     &  &   \\
				\multirow{2}{*}{60$^{\circ}$}     & III      &  $75.4\pm1.8$   & \,\;9    & \multirow{2}{*}{866}      &  $0.0871\pm0.0021$    & \multirow{2}{*}{$1.166\pm0.059$}   \\
				& I        &  $64.7\pm2.9$   &   10    &                                      &  $0.0747\pm0.0033$    &   \\
				\hline
			\end{tabular}
		}
		\label{tb:scenarios}
	\end{center}
\end{table}

\begin{figure}[!h]
	\begin{center}
		\includegraphics [width=0.65\textwidth]{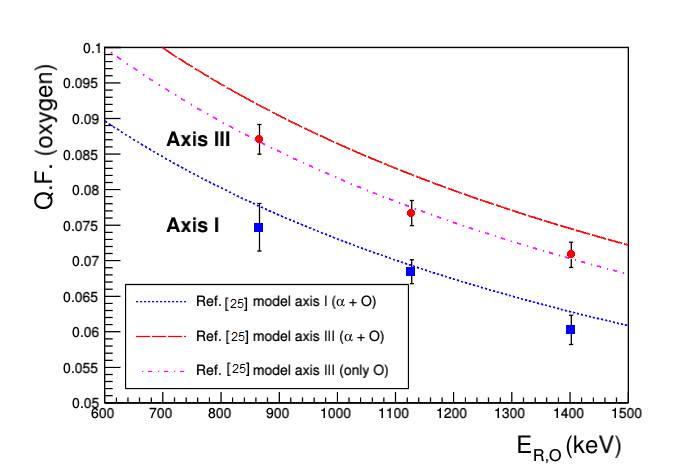}
	\end{center}
	\vspace{-0.8cm}
	\caption{\footnotesize{
			Quenching factors for oxygen nuclear recoils in ZnWO$_4$ for the crystal axes I and III as function of the expected recoil energies $E_{R,O}$.
			In the plot the expected behavior of quenching factor for the two crystal axes of the model of Ref. \cite{Tret2010} are also reported; they have been obtained by  fitting the data of the $\alpha$'s and oxygen recoil data together, and oxygen recoil data only.
	}}
	\label{fg:qf}   
\end{figure}

In Table \ref{tb:scenarios} the summary of the peak positions measured for the three considered 
angles is reported. The expected recoil energies for the oxygen nucleus, $E_{R,O}$, are also given.
The obtained quenching factors are calculated as the ratio $E/E_{R,O}$ and
the last column reports the degree of anisotropy for each scattering-angle/recoil energy.

In Fig. \ref{fg:qf} the obtained value for the quenching factors are plotted together with the models 
for the considered crystal axes derived from Ref. \cite{Tret2010}. 
This parameter has been estimated, for each axis, 
taking into account the response of ZnWO$_4$ to $\alpha$ particles and to oxygen recoils (see Ref. \cite{neut19}).

\section{Conclusions}

The development and the results obtained with a ZnWO$_4$ scintillation detector and reported here
demonstrate that this anisotropic scintillator is very promising 
to exploit the directionality technique to investigate those DM particle inducing recoils
of target nuclei.

The results confirm the anisotropic response of the ZnWO$_4$ crystal scintillator to $\alpha$ 
particles and to oxygen nuclear recoils. The presence of a good anisotropic response also in 
the lower energy region is supported by the trend of the measured quenching factors. 
This could open the possibility to realize a pioneer ADAMO experiment to investigate  
the mentioned DM candidates by means of directionality through the use of anisotropic scintillators.

\end{document}